\documentclass[aps,prl,twocolumn,superscriptaddress]{revtex4}
\newif\ifpdf
\ifx\pdfoutput\undefined
\pdffalse % we are not running PDFLaTeX
\else
\pdfoutput=1 % we are running PDFLaTeX
\pdftrue
\fi

\ifpdf
\usepackage[pdftex]{graphicx}
\else
\usepackage{graphicx}
\fi
\usepackage{hyperref}
\usepackage{amssymb}
\usepackage{epsfig}
\usepackage{amsbsy}

\begin{document}

\ifpdf
\DeclareGraphicsExtensions{.pdf, .jpg}
\else
\DeclareGraphicsExtensions{.eps, .jpg}
\fi

\def\hslash{\hbar}
\def\imag{i}
\def\grad{\vec{\nabla}}
\def\div{\vec{\nabla}\cdot}
\def\curl{\vec{\nabla}\times}
\def\DDt{\frac{d}{dt}}
\def\ddt{\frac{\partial}{\partial t}}
\def\ddx{\frac{\partial}{\partial x}}
\def\ddy{\frac{\partial}{\partial y}}
\def\lap{\nabla^{2}}
\def\divv{\vec{\nabla}\cdot\vec{v}}
\def\gradS{\vec{\nabla}S}
\def\vvec{\vec{v}}
\def\wc{\omega_{c}}
\def\<{\langle}
\def\>{\rangle}
\def\Tr{{\rm Tr}}
\def\Csch{{\rm csch}}
\def\Coth{{\rm coth}}
\def\Tanh{{\rm tanh}}
\def\g2{g^{(2)}}

% Use the \preprint command to place your local institutional report
% number in the upper righthand corner of the title page in preprint mode.
% Multiple \preprint commands are allowed.
% Use the 'preprintnumbers' class option to override journal defaults
% to display numbers if necessary
%\preprint{}

%Title of paper
\title{Exciton dissociation
at donor-acceptor polymer heterojunctions: quantum nonadiabatic dynamics
and effective-mode analysis}

% repeat the \author .. \affiliation  etc. as needed
% \email, \thanks, \homepage, \altaffiliation all apply to the current
% author. Explanatory text should go in the []'s, actual e-mail
% address or url should go in the {}'s for \email and \homepage.
% Please use the appropriate macro foreach each type of information

% \affiliation command applies to all authors since the last
% \affiliation command. The \affiliation command should follow the
% other information
% \affiliation can be followed by \email, \homepage, \thanks as well.
\author{Hiroyuki Tamura}
\affiliation{D\'epartement de Chimie, 
Ecole Normale Sup\'erieure,
24 rue Lhomond, F--75231 Paris cedex 05, France}

\author{Eric R. Bittner}
\affiliation{Department of Chemistry and Texas Center for
  Superconductivity,University of Houston, Houston, Texas 77204}
 
 \author{Irene Burghardt}
  \affiliation{D\'epartement de Chimie, 
Ecole Normale Sup\'erieure,
24 rue Lhomond, F--75231 Paris cedex 05, France}

%Collaboration name if desired (requires use of superscriptaddress
%option in \documentclass). \noaffiliation is required (may also be
%used with the \author command).
%\collaboration can be followed by \email, \homepage, \thanks as well.
%\collaboration{}
%\noaffiliation

\date{\today}

\begin{abstract}
The quantum-dynamical mechanism of photoinduced subpicosecond
exciton dissociation and the concomitant formation of a charge-separated
state at a TFB:F8BT polymer heterojunction is elucidated. The analysis
is based upon a two-state
vibronic coupling Hamiltonian including
an explicit 24-mode representation of a
phonon bath comprising high-frequency (C$=$C stretch) and 
low-frequency (torsional) modes.
The initial relaxation behavior is characterized by coherent oscillations,
along with the decay through an extended nonadiabatic coupling region.
This region is located in the vicinity of a conical intersection 
hypersurface. A central ingredient of the analysis is a novel
effective mode representation, which highlights the role of the
low-frequency modes in the nonadiabatic dynamics.
Quantum dynamical simulations were carried out using the 
multiconfiguration time-dependent Hartree (MCTDH) method. 
\end{abstract}

% insert suggested PACS numbers in braces on next line
% insert suggested keywords - APS authors don't need to do this
%\keywords{}

%\maketitle must follow title, authors, abstract, \pacs, and \keywords
\maketitle

\begin{figure}[t]
  \includegraphics[angle=0,width=\columnwidth]{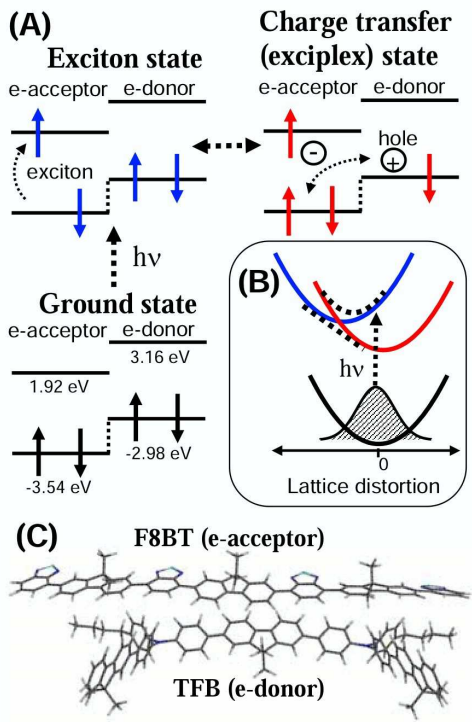}
\caption{
(A)\ (Color online) Schematic diagram illustrating 
the electronic structure and 
photophysics of the TFB:F8BT donor-acceptor 
heterojunction. 
The solid and dashed lines indicate 
the energy levels and band offset $\Delta E$, respectively; the 
solid arrows indicate the electron occupancies.
(B)\ Simplified one-dimensional scheme of 
the photoexcitation and subsequent nonadiabatic potential 
crossing process. 
The black, blue and red curves indicate 
the ground, exciton and charge transfer states, respectively. 
The dashed lines indicate the adiabatic potentials at the 
avoided crossing.
(C)
%Molecular structures of the F8BT (acceptor) and 
%TFB (donor) monomers.
Optimized molecular geometry, from density functional theory 
(DFT) calculations, of the parallel TFB and F8BT
polymer chains at the heterojunction interface. }
\end{figure}

The photophysics of $\pi$-conjugated
organic semiconductor systems is a key
ingredient in the technological development of
optoelectronic devices.\cite{Friend97}
Due to their spatially extended structure, these systems exhibit
both the molecular characteristics of their components and the
collective electronic excitations (exciton states) characteristic
of lattice structures. The coupling between electronic and nuclear
motions results in the localization (``self-trapping'') of exciton
states, along with the coherent nuclear dynamics and vibronic coupling
phenomena known from molecular
systems.\cite{SR06,BR06,RB06,BRK05,KB03a,KB03,KBB01}
Experiments have indeed
provided evidence for coherent vibrational
dynamics\cite{LCBS03} and
ultrafast (typically $\sim$ 100 fs) decay of electronic
excitations.\cite{Ketal93,Setal06}

Of particular interest are ultrafast exciton dissociation
processes at polymer heterojunctions,\cite{BR06,RB06,BRK05,Setal06,Hetal99,Metal04} 
comprising two materials with a $\Delta E$ off-set between the 
constituent valence and conduction bands (see Fig.\ 1).
In molecular terms, 
this is the energy difference between the HOMO or LUMO levels
of the two components. 
If $\Delta E$ is large compared to the exciton binding energy
$\epsilon_B$ (for typical
organic polymers, $\epsilon_B \approx 0.5$ eV),
the photoexcited exciton state decays to 
an interfacial charge-separated state (or 
exciplex).\cite{BR06,RB06,BRK05,Setal06,Hetal99,Metal04}
Generally, such a situation is desirable for
photovoltaic systems where $\Delta E$ provides the driving force for
charge-separation upon photoexcitation. 
On the other hand, if $\epsilon_B > \Delta E$, 
the excitonic state is stable, and such situations are best suited for
light-emitting diode (LED) applications.
If $\epsilon_B\simeq\Delta E$, charge separation and
exciton regeneration processes tend to compete.\cite{Setal06,Hetal99,Metal04}

%\vspace*{0.2cm}

Even though considerable progress has been achieved over recent years
in the electronic structure characterization of polymer semiconductor
systems,\cite{BR06,BRK05,Setal06}
only tentative explanations have been given for 
the quantum-dynamical mechanism
of the relevant charge transfer and decay processes.\cite{BR06,PB06}
This applies, in particular, to the electronically non\-adiabatic
decay channels which are of key importance for the ultrafast
phenomena
mentioned above. Given the presence of multiple electronic state
crossings in conjunction with electron-phonon vibronic couplings,
nonadiabatic crossing regions and
conical intersection topologies are expected to play a landmark role. 
Against this background, this Communication
aims to provide a detailed quantum-dynamical picture
of the non\-adiabatic events determining exciton dissociation and
formation of a charge transfer (exciplex) state at a 
polymer heterojunction.

In the following, we focus upon the fate of the
primary photo-excitation at a 
donor-acceptor heterojunction formed at the 
phase-boundary between 
poly[9,9-dioctylfluorene-co-N-(4,butylphenyl)diphenylamine] (TFB)
 and  poly[9,9-dioctylfluorene-co-benzothiadiazole] (F8BT) when 
 spin-cast from solution.\cite{Metal04,morteani:1708} 
 The optimized molecular geometry of the two polymer chains 
 at the interface is shown in Fig. 1.  
 LEDs fabricated using these materials exhibit considerable
 luminescence intensity. 
This  is remarkable because
the relative HOMO energy off-set between the two polymers,  $\Delta E$, is
only slightly 
larger than $\epsilon_B$.\cite{BR06,RB06,BRK05,Setal06,Metal04}
The photoexcited exciton (XT) state
is thus only marginally stable against
dissociation into a 
charge-transfer (CT) exciplex state,
and the majority of the primary exciton population
is indeed found to undergo charge separation.\cite{Metal04,morteani:1708}
Although the CT state is weakly emissive, 
appearing 140$\pm$20 meV to the red of the exciton,
the considerable luminescence is 
attributed to secondary excitons formed by endothermic back 
transfer of the hole from TFB$^+$ to F8BT$^-$ to form the F8BT exciton
with an activation energy of $100\pm 30$ meV.\cite{Metal04,morteani:1708,foot1}Nonadiabatic XT/CT state interactions clearly play a crucial role in
this system.

\begin{figure}[b]
\includegraphics[angle=0,width=\columnwidth]{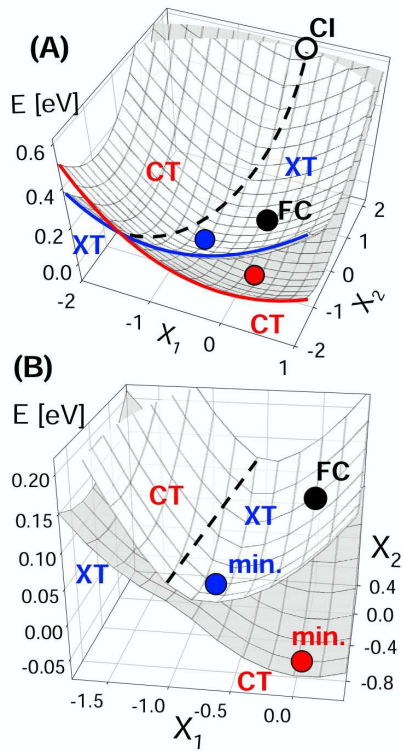}
\caption{(Color online) 
(A) Coupled adiabatic potential energy surfaces (PESs) 
as a function of the branching plane coordinates 
$(X_1,X_2)$,  
and (B) zoom-in on the avoided-crossing region. 
The black, white, blue and red circles indicate the respective 
locations of 
the Franck-Condon (FC) geometry, the conical intersection (CI),
and the minima of the exciton (XT) and charge transfer (CT) 
states, respectively.
The blue and red lines indicate the XT and CT 
diabatic states, respectively.   
The dashed line indicates the XT-CT avoided-crossing seam line. }
\end{figure}

In order to understand how nuclear motion influences the electronic
transfer and decay
mechanisms at the TFB:F8BT heterojunction, we carried out 
quantum-dynamical simulations based upon the two-state
vibronic coupling model put forward by one of us in 
Refs.\cite{BR06,PB06}.
A subpicosecond scale decay from the XT
state is induced by vibronic interactions involving high-frequency 
(C$=$C stretch) and low-frequency (torsional) modes. 
As explained below, we identify an extended nonadiabatic coupling
region in the vicinity of a multi-dimensional intersection
space. While the electron-phonon coupling is largely carried by
the high-frequency modes, the low-frequency modes are shown to
play a key role in the decay dynamics. The main features of this
dynamics are expected to carry over to a general class of conjugated
polymer systems.

%\vspace*{0.2cm}

The following analysis makes extensive use of a recently developed
effective-mode description of the nonadiabatic dynamics at conical
intersections,\cite{CGB05,BGC06,GBC06a,GBC06b,B06} in 
conjunction with efficient multiconfigurational
quantum propagation techniques.\cite{MMC90,MMC92,BJWM00,MCTDH_HD} 
Our previous analysis has shown that the cumulative effects of
many modes in a nonadiabatic coupling situation
can be represented by {\em three effective modes}
that entirely determine the short-time dynamics.
In the present study, we further connect this previous analysis to
a systematic, Mori-chain type decomposition of the
phonon bath\cite{TB06,GP82} (see also the related development in 
Ref.\cite{GC06}). 
As shown below, the explicit 24-mode phonon bath of the model
under discussion can thus be
replaced by a 9-mode effective bath over the time scale 
of interest.

%\vspace*{0.2cm}

Our starting point is the following electron-phonon
Hamiltonian for the coupled XT and CT 
states interacting with an $N$-mode phonon bath,\cite{PB06,KDC84} 
\begin{eqnarray}
  \label{HLVC}
  \boldsymbol{H} & = &
  \boldsymbol{V}_\Delta +
  \sum_{i=1}^N \boldsymbol{H}_i 
\nonumber 
\end{eqnarray}

%\vspace*{-0.4cm}

where 

%\vspace*{-0.5cm}

\begin{eqnarray}
  \label{HLVC_2}
  \boldsymbol{H}_i = \frac{\omega_i}{2}
  \biggl( p_i^2\, +\, x_i^2 \biggr) {\bf 1} 
  + \left( \begin{array}{cc}
  {\kappa}_i^{(1)}\, x_i  &  \lambda_i x_i  \\
  \\
  \lambda_i x_i        & {\kappa}_i^{(2)}\, x_i
  \end{array} \right) 
\end{eqnarray}
with $p_i^2 = -\partial^2/\partial x_i^2$ and the electronic 
splitting $\boldsymbol{V}_\Delta = - \Delta \boldsymbol{\sigma}_z$
(with $\boldsymbol{\sigma}_z$ the Pauli matrix). 
Mass and frequency weighted coordinates were used, along with the 
convention $\hbar=1$. As mentioned above, the
phonon bath comprises a high-frequency 
branch composed of C$=$C stretch modes and a low-frequency branch
of torsional modes. An explicit representation in terms of $N$=24
modes was introduced according to Ref.~\cite{PB06}. 

%\vspace*{0.2cm} 

Eq.\ (\ref{HLVC}) corresponds to a general linear vibronic coupling 
form, in a diabatic representation,\cite{KDC84}
and was parameterized by semiempirical calculations 
as described in Refs.\cite{BR06,BRK05,KB03}. The
Hamiltonian Eq.\ (\ref{HLVC}) gives rise
to an $(N-2)$-dimensional conical intersection
space,\cite{KDC84} which is the 
key feature determining the nonadiabatic dynamics of the system. 
Indeed, the nonadiabatic decay induced by conical intersections 
is often ultrafast, i.e., occurring on a subpicosecond to picosecond 
time scale. 

%\vspace*{0.2cm} 

For the purpose of the following discussion, a 
decomposition of the phonon bath in terms of collective, or effective 
modes is introduced. To this end, we use an orthogonal
coordinate transformation, 
$\boldsymbol{X} = \boldsymbol{T} \boldsymbol{x}$,\cite{CGB05,BGC06,GBC06a,GBC06b} leading to 
the following decomposition in terms of 
{\em effective} vs.\ {\em residual} modes, 
\begin{eqnarray}
  \label{Heff_res}
\boldsymbol{H} = \boldsymbol{H}_{\rm eff} + \boldsymbol{H}_{\rm res}
\end{eqnarray}

\begin{widetext}
with the part $\boldsymbol{H}_{\rm eff}$ which contains {\em three
effective modes} $(X_1,X_2,X_3)$ that entirely define the coupling 
to the electronic subsystem,
\begin{eqnarray}
  \label{Heff}
  \boldsymbol{H}_{\rm eff} & = &  
  \boldsymbol{V}_\Delta +
  \sum_{i=1}^{3}
  \frac{\Omega_i}{2} ( P_i^2\, +\, X_i^2 )\ {\bf 1}
  + \sum_{i,j=1,j>i}^{3}
  d_{ij} ( P_i P_j + X_i X_j )\ {\bf 1}
  \nonumber \\
  \nonumber \\  
  \mbox{} &  & \hspace*{1.0cm}  + \sum_{i=1}^{3} K_i X_i {\bf 1} + 
  \left( \begin{array}{cc}
  D_1 X_1 + D_2 X_2 &  \Lambda X_1  \\
  \\
  \Lambda X_1   &  - D_1 X_1 - D_2 X_2
  \end{array} \right) 
\end{eqnarray}
Here, a topology-adapted representation\cite{BGC06} was chosen, where 
$(X_1,X_2)$ lift the degeneracy at the intersection
(and thus span the branching plane\cite{AXR91}), while $X_3$ 
lies in the intersection space. The modes $(X_1,X_2)$ define a suitable
reduced representation of the intersecting surfaces, as illustrated in 
Fig.\ 2. The parameters of Eq.\ (\ref{Heff}) relate to
the original Hamiltonian Eq.\ (\ref{HLVC}) as described in 
Refs.\cite{BGC06,TB06}. 

\end{widetext}
%\vspace*{0.2cm} 

The residual
Hamiltonian $\boldsymbol{H}_{\rm res}$ contains the 
remaining $(N-3)$ modes, and their
bilinear coupling to the effective modes
and among each other, 
\begin{eqnarray}
     \label{Hres}
     \boldsymbol{H}_{\rm res} &=& \sum_{i=4}^{N}
     \frac{\Omega_i}{2} (P_i^2 + X_i^2) {\bf 1}  \nonumber \\
     &+& \sum_{i=1}^N \sum_{j=4}^{N}
     d_{ij} \biggl(
     P_i P_j + X_i X_j \biggr)
     {\bf 1}
\end{eqnarray}

%\vspace*{0.2cm}

Importantly, $\boldsymbol{H}_{\rm res}$ is diagonal 
with respect to the electronic subspace.

%\vspace*{0.2cm} 

The transformation leading to Eqs.\ (\ref{Heff_res})-(\ref{Hres})
introduces a hierarchical structure in the phonon bath: While the
effective modes 
couple directly to the electronic two-level system, the 
residual modes couple in turn to the effective modes. The
new choice of coordinates has important implications for possible
approximations, since the effective mode Hamiltonian 
$\boldsymbol{H}_{\rm eff}$
by itself preserves 
the first three moments of the overall Hamiltonian.\cite{GBC06a} 
Hence, the 
{\em short-time dynamics is entirely described 
by $\boldsymbol{H}_{\rm eff}$}.

%\vspace*{0.2cm}

The concept of a mode hierarchy can be carried further by 
introducing additional transformations within the 
subspace of residual modes. In particular, 
$\boldsymbol{H}_{\rm res}$ can be transformed to a 
band-diagonal form,\cite{TB06} corresponding to a 
{\em hierarchy of residual bath modes} (see also Ref.\cite{GC06}
for a related analysis). An approximate $n$th-order,
i.e., $(3 + 3n)$-mode Hamiltonian can be defined 
by truncating this hierarchy at a given order, 
\begin{eqnarray}
\label{Hn}
\boldsymbol{H}^{(n)} = \boldsymbol{H}_{\rm eff} + 
\sum_{l=1}^{n}\, {\boldsymbol{H}_{\rm res}^{(l)}}
\end{eqnarray}
with the $l$th order residual bath Hamiltonian 
\begin{eqnarray}
     \label{Hres_n}
     \boldsymbol{H}_{\rm res}^{(l)} &=& \sum_{i=3l+1}^{3l+3}
     \frac{\Omega_i}{2} (P_i^2 + X_i^2) {\bf 1}  \nonumber \\
     &+& \sum_{i=3l+1}^{3l+3} \sum_{j=i-3}^{i-1}
     d_{ij} \biggl(
     P_i P_j + X_i X_j \biggr)
     {\bf 1}
\end{eqnarray}

%\vspace*{0.2cm}

The results shown in Fig.\ 3, to be discussed in detail below, 
illustrate the effect of truncating the hierarchy of residual bath modes. 

%\vspace*{0.2cm} 

The scheme of Eqs.\ (\ref{Hn})-(\ref{Hres_n}) corresponds 
to a generalized Mori chain, which can be shown to conserve 
successive orders of the 
Hamiltonian moments.\cite{GP82} 
If truncated at a given order, the hierarchy can be formally 
closed by adding Markovian dissipation.\cite{TB06,GP82} 

%\vspace*{0.2cm}

As can be inferred from the reduced representation of Fig.\ 2,
in the ($X_1,X_2$) subspace, a conical intersection indeed
exists in the present system, and an extended avoided crossing region
is present in its vicinity. 
The figure further illustrates that the minima of both
electronic states are located on the same side of an
avoided-crossing seam that departs from the conical intersection.
Following the terminology of Marcus
theory, the system can thus be classified in terms of the ``inverted
regime''.\cite{BR06} Due to the prevalent weak coupling in the
avoided-crossing region, the zeroth-order picture of the dynamics
is expected to be markedly diabatic.
With a Franck-Condon (FC) initial condition ($X_1 = X_2 = 0$), the
wavepacket rapidly accesses the nonadiabatic coupling 
region, but does not directly encounter the conical intersection. 

%\vspace*{0.2cm}

Fig.\ 3 (trace ``exact'') shows results of
quantum-dynamical calculations for the
overall 24-mode system according to the Hamiltonian Eq.\ (1), using
the multiconfiguration time-dependent Hartree (MCTDH)
method.\cite{MMC90,MMC92,BJWM00,MCTDH_HD} From the figure, one can infer
that the XT state population has decayed to about 
50\% after 200 fs, with a
corresponding increase in the CT state population.
Subsequent to the initial decay, the populations
of the two states remain approximately equal, with an oscillatory
behavior which features the characteristic period of the
slow (torsional) modes ($T_{\rm torsion} \sim 300$ fs). Superimposed
is a weak oscillatory structure due to the fast (C$=$C stretch) modes
($T_{\rm stretch} \sim 20$ fs). The dynamics apparently remains in a
coherent regime over the observation period, and does not reach an
equilibrium state.

%\vspace*{0.2cm}

The effective mode analysis as outlined above is
expected to shed further light on the mechanism of the nonadiabatic
decay of the XT state. Fig.\ 3 illustrates the results of wavepacket
propagation for the successive orders $n = 0, \ldots, 2$ (i.e.,
3 to 9 effective phonon modes) of the effective-mode hierarchy.
These successive orders
feature in alternation the high-frequency
modes (for $n=0$ and $n=2$) and the low-frequency modes (for
$n=1$).\cite{TB06}

%\vspace*{0.2cm}

Several observations can be inferred from Fig.\ 3: (i) At the
order $n = 0$ (3 effective modes), i.e.,
for $\boldsymbol{H} = \boldsymbol{H}_{\rm eff}$,
no decay of the exciton state is observed. While the effective-mode
construction guarantees that the very initial dynamics is correctly
reproduced,\cite{GBC06a}
deviations from the exact dynamics occur from about 80 fs onwards.
(ii) At the order $n=1$ (6 effective modes), the XT $\rightarrow$ CT
decay over the first 400 fs is correctly described. (iii) At the order
$n=2$ (9 effective modes), the dynamics is correctly reproduced over the
entire time scale of observation (1.5 ps).

\begin{figure}[h]
\includegraphics[angle=0,width=\columnwidth]{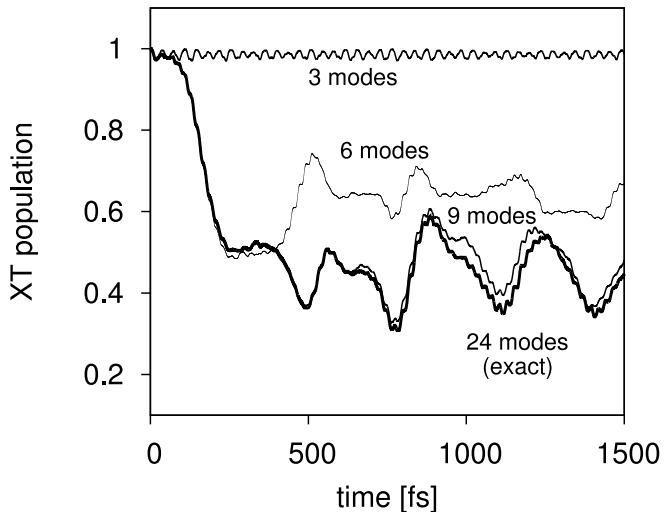}
\caption{
Time-dependent population of the exciton (XT) 
diabatic state, from MCTDH calculations for the 
overall 24-mode system (bold line, labeled ``exact'') 
and the successive 
effective-mode approximations 
$\boldsymbol{H}^{(0)} = \boldsymbol{H}_{\rm eff}$ (3 modes), 
$\boldsymbol{H}^{(1)}$ (6 modes), and $\boldsymbol{H}^{(2)}$ (9 modes),
see Eqs.\ (5)-(6). 
Note that the 9-mode result is very close to the exact 
24-mode calculation.}
\end{figure}

Strikingly, the
presence of the high-frequency modes alone, at the 
level of the $n = 0$ description, does not account for 
the XT $\rightarrow$ CT transition. A more detailed analysis 
shows that the dynamics is characterized by oscillations between the
adiabatic states, but the population essentially remains confined
to the XT branch. The overall picture in the reduced 3-mode space
is thus predominantly diabatic. 

%\vspace*{0.2cm}

At the order $n = 1$ (6 effective modes), the low-frequency
(torsional) branch of the phonon bath is added. As illustrated 
in Fig.\ 3, the initial XT state decay is now reproduced correctly.
Apparently the low-frequency modes play a key role in the
nonadiabatic dynamics, even though they do not directly couple
to the electronic subsystem in the effective-mode picture.
A study of the energy flow among the effective modes
reveals that the low frequency modes act so as to dissipate
the energy contained in the 3-mode $(X_1,X_2,X_3)$ subspace,
by an intramolecular vibrational redistribution (IVR) process.\cite{TB06}
In addition, the non-adiabatic dynamics as such is
influenced by the low-frequency components.

%\vspace*{0.2cm}

The next higher level,
$n = 2$ (9 effective modes), yields very good agreement with the
24-mode reference calculation over the whole propagation period.
The $\boldsymbol{H}^{(2)}$ approximation thus provides a suitable
surrogate Hamiltonian on the time scale of observation.

%\vspace*{0.2cm}

Finally, Fig.\ 4
shows the time-dependent $(X_1,X_2)$ position expectation values for the
XT vs.\ CT portions of the wavepacket, from the 9-mode calculation. 
A concerted, oscillatory motion persists over the complete 
observation period. The comparatively regular behavior
is consistent with repeated passages through a weakly
avoided crossing region, rather than a direct passage through a conical
intersection. The presence of the conical intersection space is
essential, though, in that it defines the relevant nonadiabatic
coupling region.

%\vspace*{0.2cm}

In summary, the present study emphasizes the role of coherent,
quantum
dynamical evolution, and its intrinsically multi-dimensional nature,
in characterizing the nonadiabatic decay of the photochemically
accessed XT state.
Note that the presence of
a high-dimensional intersection topology is not an exception, but
rather the rule in multidimensional systems.\cite{KDC84,TM03} 
The effective-mode analysis proposed here leads to 
a reduced dimensionality description, which provides
important insight 
both from a static and a dynamical viewpoint. The representation 
of the potential in Fig.\ 2, in terms
of the effective modes $(X_1,X_2)$, yields a unique picture of
the topology of the coupled electron-phonon system.
From a dynamical point of view, the effective modes
$(X_1,X_2,X_3)$ determine the shortest, initial time scale, 
i.e., several tens of femtoseconds. For the present system, it is
however essential to (at least partially) include the residual
$(N-3)$-mode phonon bath. As demonstrated in Fig.\ 3, a 9-mode
representation yields a very good approximation of
the overall 24-mode dynamics.

\begin{figure}
\hspace*{0.5cm} \includegraphics[angle=0,width=\columnwidth]{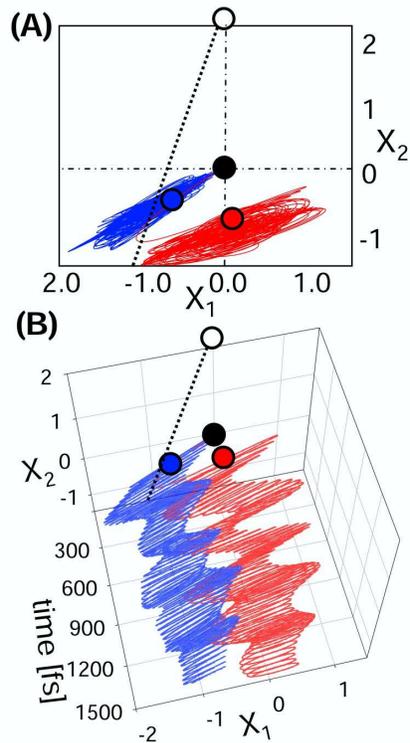}
\caption{
(Color online) 
Time-dependent position expectation values 
of the XT (blue) and CT (red) portions of the wave-packet 
as a function of the branching plane coordinates
$(X_1,X_2)$, for the $n = 9$ calculation.
The black, white, blue and red circles indicate the locations of 
the FC geometry, the conical intersection, and the 
XT vs.\ CT minima, respectively.
The dashed line indicates the XT-CT avoided-crossing seam line.
(A) Projection on the $(X_1,X_2)$ plane and (B) 
3-dimensional $(X_1,X_2,t)$ plot. }
\end{figure}

%\vspace*{0.2cm}

One of the main conclusions of this study is that the
presence of the low-frequency (torsional) modes is a key element
in the observed decay dynamics. Even though the high-frequency
(C$=$C stretch)
modes are most strongly coupled to the electronic subsystem,
they do not by themselves induce the observed XT state decay.
The coupling to the low-frequency phonon branch 
gives rise to an energy redistribution (IVR) which is essential in
mediating the nonadiabatic decay. 
The importance
of the low-frequency branch has been previously recognized in
the context of the static absorption/emission spectroscopy of
similar systems.\cite{KB03,KBBM00}

%\vspace*{0.2cm}

Our observations are in qualitative
agreement with experimental results providing evidence for
exciton regeneration at the TFB:F8BT 
heterojunction.\cite{BR06,Metal04} Even though our calculations
do not account for temperature effects, and the influence of
dissipation is limited to the explicitly included phonon modes, we expect
that the coherent, non-equilibrium nature of the dynamics remains a
dominant feature over the first few picoseconds.
Further studies including temperature effects will allow for a
detailed comparison with the results, e.g., of the non-Markovian master
equation calculations reported in Ref.\ [\cite{PB06}].

\begin{acknowledgments}
We would like to thank Andrey Pereverzev, Etienne Gindensperger, and 
Lorenz Cederbaum for constructive discussions. 
This work was supported by the ANR-05-NANO-051-02 project, by NSF grant
CHE-0345324, and by the Robert Welch Foundation.
\end{acknowledgments}


\begin{thebibliography}{10}
\bibitem{foot1}
As a result, the 
overall photoluminescence efficiency of the TFB:F8BT blend is  
$\approx 60 \%$ that of the pure F8BT, in spite of the presence of a 
heterojunction.
\end{thebibliography}
\end{document}